# Thermodynamic Entropy And The Accessible States of Some Simple Systems

David Sands
Dept. of Physical Sciences
University of Hull
Hull
HU6 7RX

e-mail: d.sands@hull.ac.uk

**Abstract**
Comparison of the thermodynamic entropy with Boltzmann's principle shows that under conditions of constant volume the total number of arrangements in simple thermodynamic systems with temperature-independent heat capacities is $T^{C/k}$. A physical interpretation of this function is given for three such systems; an ideal monatomic gas, an ideal gas of diatomic molecules with rotational motion, and a solid in the Dulong-Petit limit of high temperature. $T^{1/2}$ emerges as a natural measure of the number of accessible states for a single particle in one dimension. Extension to N particles in three dimensions leads to $T^{C/k}$ as the total number of possible arrangements or microstates. The different microstates of the system are thus shown *a posteriori* to be equally probable, with probability $T^{-C/k}$, which implies that for the purposes of counting states the particles of the gas are distinguishable. The most probable energy state of the system is determined by the degeneracy of the microstates.

**Introduction**
Entropy is one of the concepts that undergraduate physicists experience most difficulty in understanding. Thermodynamically, only changes in entropy are important and though these can be defined and calculated very easily, there is no easy answer as to what exactly entropy itself is. Statistical mechanics has put a probabilistic interpretation on the quantity, but there is a real conceptual difficulty in so far as the Boltzmann entropy does not always match the thermodynamic entropy. Jaynes has shown, for example, that whereas the Gibbs entropy is numerically equal to the thermodynamic entropy, the Boltzmann entropy is generally smaller in magnitude unless the gas is ideal [1].

The essential problem with the statistical view lies in the concept of disorder first invoked by Boltzmann and associated with entropy ever since. As Jaynes has been keen to stress, the association of disorder with entropy is a misconception. The very word "disorder" implies an irregular or unstructured arrangement, but the use of probability distributions arises from a lack of knowledge about the system. No matter how apparently unstructured or random a given arrangement might seem, if the arrangement could be determined in some way there would be no need to resort to probability to describe the system. Entropy is therefore anthropomorphic; it simply depends on your point of view [2].

The connection between thermodynamic entropy and statistical entropy therefore seems remote. Thermodynamically entropy appears to be absolute, in so far as the third law requires it to go to zero at zero Kelvin, but statistical entropy seems to be relative. It is the purpose of this paper to show that in fact there is an intuitively accessible connection between thermodynamics and statistics because the thermodynamic entropy itself implies a statistical distribution via the function $W=T^{C/k}$, derived as follows. Thermodynamically, a small change in heat dQ produces a change in entropy dS according to

$$dS = \frac{dQ}{T} \qquad 1$$

For a system with constant heat capacity between the temperatures $T_1$ and $T_2$, such as an ideal gas at constant volume or a solid in the Dulong-Petit limit, then

$$\Delta S = c_v \ln(T_2) - c_v \ln(T_1) \qquad 2$$

This can be compared with Boltzmann's entropy, defined as the logarithm of the number of microscopic arrangements W.

$$S = k \ln(W) \qquad 3$$

The change in entropy consequent upon heat input is therefore

$$\Delta S = S_2 - S_1 = k \ln(W_2) - k \ln(W_1) \qquad 4$$

It follows that

$$W = T^{c_v/k} \qquad 5$$

The derivation of (5) is straightforward and standard, but to the author's knowledge no physical interpretation has been placed upon it. There is, of course, an immediate mathematical interpretation. If $kT$ represents a basic unit of thermal energy and $c_vT$ represents the heat that system would contain if the heat capacity were constant over the entire range of temperatures, then $c_v/k$ represents the number of thermal units. $c_v/k$ could be regarded as the number of micro-capacitors, each of thermal capacity $k$, over which the energy is distributed. The average energy per micro-capacitor would be $kT$, but if each were considered to have a range of $T$ possible values of energy there would in fact be $T^{C/k}$ possible permutations. Mathematically this is perfectly permissible, but it begs the obvious question as to why should there be $T$ possible values of energy.

In fact it will be shown that the number of possible values is not $T$ but $T^{1/2}$. That is, each particle can be in one of $T^{m/2}$ states in one dimension, where m is the number of modes of motion. Extended to $N$ particles in three dimensions, this leads naturally to equation (5) and hence to a simple microscopic interpretation of the thermodynamic entropy.

**An ideal monatomic gas**
Consider first a monatomic gas consisting of $N$ atoms. As is well known, the heat capacity is $c_v=3/2.k.N$. The distribution of molecular speeds is given by the Maxwell-Boltzmann speed distribution which becomes a Gaussian in one dimension[3] with standard deviation

$$\sigma = \frac{v_m}{\sqrt{2}} \qquad 6$$

Here $v_m$ is the most probable velocity

$$kT = \frac{mv_m^2}{2} \qquad 7$$

A Gaussian distribution of velocities is of course continuous so it is necessary therefore to define an interval $\delta v$ with a corresponding probability $p(v).\delta v$ of finding a particle with a velocity in the range $v$ to $v+\delta v$.

Let the total range of the Gaussian distribution be $\lambda\sigma$, where $\lambda$ is an integer. The properties of a Gaussian distribution dictate that over 99% of the velocities will lie in the range defined by $\lambda=6$ but in fact the value need not be specified exactly. Let the velocity interval be some fraction of $\lambda$, say $\delta v=\alpha\lambda$. If there are a total of $\gamma$ intervals over the whole range of $v$, then

$$\lambda\sigma = \gamma\delta v = \gamma\alpha.\lambda \qquad 8$$

Therefore

$$\gamma\alpha = \sqrt{\frac{kT}{m}} \qquad 9$$

As $k$ and $m$ are constants of the system, it is possible to identify $\alpha=\sqrt{(k/m)}$ and $\gamma=\sqrt{T}$. In one dimension then, a molecule can lie in any one of $T^{1/2}$ possible ranges of velocity, and as each velocity component is independent of the other there are $T^{3/2}$ possibilities taken over the three dimensions. Hence for $N$ molecules there are $T^{3N/2}$ possibilities, as required by the thermodynamic entropy.

The preceding treatment has concentrated on the accessible states of the system, but if the interpretation is valid it should be possible to reconstruct the entropy from the probability distributions. If the probability of a particle having a velocity v is

$$p(v) = p(v_x).p(v_y).p(v_z) \qquad 10$$

then the entropy is given by

$$\frac{S}{k} = -\int p(v).\ln[p(v)].dv \qquad 11$$

Resolving the probability distribution into its orthogonal terms yields

$$\int p(v).\ln[p(v)].dv = \iiint p(v_x).p(v_y).p(v_z)[\ln p(v_x) + \ln p(v_y) + \ln p(v_z)]dv_x dv_y dv_z$$

$$12$$

The x, y and z components are entirely independent so taking the first of these terms as an example;

$$\iiint p(v_x).p(v_y).p(v_z)\ln p(v_x)dv_x dv_y dv_z = \int p(v_x)\ln p(v_x)dv_x \int p(v_y)dv_y \int p(v_z)dv_z$$

$$13$$

Therefore

$$\frac{S}{k} = -\int p(v_x).\ln p(v_x)dv_x - \int p(v_y)\ln p(v_y)dv_y - \int p(v_z)\ln p(v_z)dv_z \qquad 14$$

It is easily shown that

$$-p(v_x).\ln p(v_x)dv_x = \frac{p(v_x)}{2}[\ln(2\pi) + \ln(\frac{kT}{m})] + p(v_x)\frac{mv_x^2}{2kT} \qquad 15$$

and

$$-\int p(v_x).\ln p(v_x)dv_x = \frac{1}{2}[\ln(2\pi) + \ln(\frac{kT}{m})] + \frac{\overline{mv_x^2}}{2kT} \qquad 16$$

Therefore

$$\frac{S}{k} = \frac{3}{2}\ln(T) + \frac{3}{2}\ln(\frac{2\pi k}{m}) + \frac{m}{2kT}(\overline{v_x^2} + \overline{v_y^2} + \overline{v_z^2})$$ 17

The factor of *3/2* arises from *ln[√T]* taken over the three dimensions, in complete agreement with the conceptual picture developed above. Finally, then, recognising that the last term is simply the ratio of average kinetic to thermal energy, for a gas of *N* particles

$$S = \frac{3Nk}{2}\ln(T) + \frac{3Nk}{2}\ln(\frac{2\pi k}{m})] + \frac{\langle E \rangle}{T}$$ 18

As required, the entropy is, within an additive constant, equal to $c_v lnT$.

**Rotational modes in diatomic molecules**
A similar argument can be applied to the rotational energy of simple diatomic molecules. Hydrogen, for example, exhibits monatomic behaviour at low temperatures because the rotational and vibrational motion is frozen out [4], but at higher temperatures the heat capacity increases to *5/2kN* upon excitation of rotational motion. The classical view of a linear rotating molecule allows for three modes of rotation, but as the rotation about the molecular axis does not contribute to the specific heat [5], this leaves two characteristic modes of rotation and four degrees of freedom each contributing ½kT per particle to the specific heat. In order to put a microscopic interpretation on this, however, it is necessary to specify the probability that an individual atom is rotating in one mode or another.

In the modern view the rotational modes are quantised. If it is assumed that quanta are transferred betweens molecules by collisions, the rate of arrival of the quanta at a molecule is simply given by the Poisson distribution. It is well known that for a mean greater than ~10 the Poisson distribution approximates to normality, so if the temperature is high enough, such that the mean rate of arrival is at least 10 quanta per atom per second, then the probability that the number of quanta arriving at an atom will lie between n and n+δn is therefore Gaussian and the treatment can proceed as before. If the mean number of quanta per atom is $n_{q1}$ and $n_{q2}$ for the two modes then the total rotational energy is

$$E_{rot} = N(n_{q1}\hbar\omega_1 + n_{q2}\hbar\omega_2) \approx W(T_{rot})Nk(T - T_{rot})$$ 19

Here $T_{rot}$ is the temperature at which the rotational quanta are activated and *W* is a step function. In reality the rotational degrees of freedom are not switched on as abruptly as a step function would imply (figure 1), but equation (19) serves as a good approximation provided $T \gg T_{rot}$ and the heat capacity is constant. Thus $n_{q1}$ and $n_{q2}$ are both proportional to the temperature as each contributes ½Nk to the heat capacity. The property of the Poisson distribution that the mean and the variance are equal therefore allows

$$\sigma^2 = \eta(T - T_{rot}) \approx \eta T \qquad 20$$

for large values of $T$. Here $\eta$ is some constant. It follows then that $\sigma$ varies with $T^{1/2}$ and as before the continuous probability distribution can effectively be divided into $T^{1/2}$ intervals. Thus the number of quanta per mode per atom can lie in any one of $T^{1/2}$ ranges, or equivalently an atom can exist in any one of $T^{1/2}$ states. Over all $N$ atoms and both modes there are $T^N$ possibilities. Hence, for a gas with rotational as well as translational energy the total number of permutations is $W = T^{3/2N} \cdot T^N = T^{5/2N}$.

As with the translational motion, the entropy can be derived from the probability distribution. The two quantised modes can be regarded as entirely independent of each other to the extent that the probability that n modes of quantum 1 and m modes of quantum 2 arrive at a molecule is simply the joint probability $p_1(n).p_2(m)$. As above, the entropy reduces to a sum of the entropies associated with each distribution.

$$\frac{S}{k} = -\int p_1(n) . \ln p_1(n) dn - \int p_2(m) \ln p_2(m) dm \qquad 21$$

Under the assumption of normality,

$$-p(n) \ln p(n) = p(n)[\frac{1}{2}\ln(2\pi) + \ln \sigma + \frac{(n-\bar{n})^2}{2\sigma^2}] \qquad 22$$

Expanding the last term and integrating,

$$\frac{1}{2\sigma^2}\int p(n)[n^2 - 2n.\bar{n} + (\bar{n})^2]dn = \frac{1}{2\sigma^2}[\overline{n^2} - (\bar{n})^2] = \frac{1}{2} \qquad 23$$

Each mode therefore contributes an entropy

$$\frac{S}{k} = \frac{1}{2}[\ln(2\pi)+1] + \ln \sigma \qquad 24$$

Therefore, from (20) and taking into account the two rotational modes and $N$ particles, the entropy becomes, within an additive constant,

$$S = Nk \ln T \qquad 25$$

**A crystalline solid**
The idea of quantised modes can also be found in the modern theory of the heat capacity of a solid, which is framed entirely in terms of the total number of vibrational modes. However, if the ideas developed above are to be applied here also, it is necessary consider the distribution of energy among the atoms of the solid. Unlike the M-B distribution for a gas, there is no simple distribution function for the energy of the constituent atoms so it is necessary once again to resort to the Poisson distribution to describe the rate of arrival of a phonon at a given atom. As with the rotational energy, the mean number of phonons will be proportional to the temperature. By the same argument $T^{1/2}$ is a measure of the width of the Gaussian distribution of phonons. The difficulty now is to find the correct model for the vibrational modes. A 3-dimensional solid containing $N$ atoms is equivalent to a $3N$ linear oscillators, so the number of possible permutations is $T^{3N/2}$ per mode of vibration. As with rotational energy of linear molecules, if there is more than one vibrational mode the number of permutations will increase. The Dulong-Petit law requires there to be two modes resulting in $T^{3N}$ possibilities, but the choice of model for the specific heat determines how many modes may be excited.

The Debye model is the best known formulation of the heat capacity. It owes its success to its ability to explain the $T^3$ behaviour of the heat capacity at low temperatures, but uses a continuous spectrum of modes up to some maximum frequency. Such a continuous spectrum does not sit easily in this statistical view of entropy. At high temperatures, where the Dulong-Petit law requires the heat capacity to approach $3kN$, there is in fact nothing unique about the Debye model. The Einstein model, which preceded Debye's model, has only one characteristic vibrational mode, but the heat capacity still tends to $3kN$ at high temperatures. As far as heat capacity goes, the mode spectrum at high temperature is not too important, but the requirements of the entropy are definite. A single Einstein mode does not yield the correct entropy. It is fortunate, therefore, that both the Einstein and Debye models are undoubtedly wrong[6]. The failure of the Einstein model at low temperatures is well known, the inadequacies of the Debye model less so. The Debye model owes its popularity to its use of a single parameter, the Debye temperature, but the fact that this parameter varies with temperature means that it does not provide an accurate description of the underlying physical phenomena[6].

The are other theories of the heat capacity of solids, such as those due to Born and von Karman as well as Raman, the lattice theory, and the Nernst-Lindemann theory [6]. Raman's theory seems to be an artificial construction with little theoretical justification, and besides has too many modes of vibration. The lattice theory is a calculational model of the modes of vibration derived from knowledge of the properties of the lattice. There appear to be three modes of vibration, but two are dominant whilst the third plays only a small part. These two modes will increase the total number of permutations to $T^{3N}$. The Nernst-Lindemann theory, which is entirely empirical, also has two modes. This theory, like Raman's, appears to have little or no theoretical justification, but it is known to describe heat capacities very well and has recently been given greater mathematical justification [7]. This well-known theory is superior to both the Debye model and the Einstein model in its ability to describe the temperature dependence of the heat capacity, but crucially provides exactly the right number of modes to increase the number of permutations to $T^{3N}$. Should it be desired, the entropy can be derived following the procedure in equations (21) to (25).

**Conclusion**
A simple statistical interpretation has been developed of the thermodynamic entropy of three sample systems; an ideal gas, a monatomic gas with rotational energy, and a solid in the Dulong-Petit limit of constant heat capacity. Comparison of the Boltzmann entropy with the thermodynamic entropy shows that the number of possible permutations is a very simple function of temperature and heat capacity, $T^{C/k}$, with a correspondingly simple physical interpretation based on the idea of temperature as a measure of statistical variability. This is an alternative meaning of temperature from that normally taught to undergraduates, which is based either on the zeroth law of thermodynamics or on kinetic theory. The kinetic temperature is defined with respect to the most probable speed of the Maxwell-Boltzmann distribution in (7) above and is most commonly associated with energy content: if heat is supplied so that the temperature rises the most probable speed increases. The present analysis suggests that $T$, or strictly $T^{1/2}$, should also be regarded as a measure of the number of accessible states in a system. Thus, the higher the temperature the greater variability in the distribution of the energy, as defined by the standard deviation of the 1-D MB distribution.

A temperature-independent heat capacity is a pre-requisite for the derivation of $T^{C/k}$ as the number of permutations (equations 1-4). It has to be recognised, however, that this is not essential for the interpretation, but the extension to cases of temperature-dependent heat capacity has not been discussed here. Rather, attention has focussed on the physical interpretation of $T^{C/k}$. It is immediately apparent that the function $T^{C/k}$ describes a distribution in which there are no degenerate states. Each permutation is equally likely as the next, which implies not only that the particles should be counted as distinguishable, but also that the velocities should be taken instead of speeds. There are six different ways in which three orthogonal components of velocity can be arranged, not to mention the different ways a given speed can be obtained from different combinations of orthogonal components. The function $T^{C/k}$ contains all these possibilities. This statistical interpretation of the thermodynamic entropy is not just an integration of the particle distribution, but represents the complete set of possibilities for the microscopic distribution of energy in a thermodynamic system. This applies as much to rotational and vibrational energy as it does to translational.

The probability of the system being in any one of the microscopic states is thus a uniform $T^{C/k}$. The idea of uniform probability is well known in statistical mechanics, but it is usually stated as an *a priori* assumption. Here it is derived *a posteriori* through the physical meaning attached to $T^{C/k}$. Many of the microscopic configurations are of course degenerate in energy and the states of the highest degeneracy will be those that are most observed. These will be normally distributed about the mean energy of the system[8]. Thus states of very low or very high energy are possible, but highly improbable and the present interpretation of the function $T^{C/k}$ makes this point clear in a way accessible to undergraduates; each of the possible distributions is sampled with a uniform probability $T^{C/k}$ but macroscopically the most probable energy state corresponds to the microstates with the highest degeneracy.

**References**

[1]     E T Jaynes, *Gibbs vs Boltzmann Entropies*, Am. J. Phys. 33 (1965) p391- 398

[2]     Peter Clark, *Matter, Motion and Irreversibility*, Brit. J. Phil. Sci. **33** (1982) 165-208

[3]     Francis W Sears and Gerhard L Salinger, *Thermodynamics, Kinetic Theory, and Statistical Mechanics*, 3$^{rd}$ Edition, Addison-Wesley Publishing Co. Singapore 1980

[4]     Mark W Zemansky, *Heat and Thermodynamics*, 5$^{th}$ Edition, McGraw-Hill Kogakusha Ltd, Tokyo, 1957

[5]     Arnold Sommerfeld, *Thermodynamics and Statistical Mechanics; Lectures on Theoretical Physics Vol 5*, Academic Press, New York 1967

[6]     M Blackman, *The theory of the specific heat of solids*, Rep. Prog. Phys. **8** (1941) 11-30

[7]     F E Irons, *New method for reducing the general formula for lattice specific heat to the Einstein and Nernst-Lindemann approximations*, Can. J. Phys. **81** (2003) 1015-1036

[8]     B.H.Lavenda; *Statistical Physics: A Probabilistic Approach*, (Wiley, New York) 1991

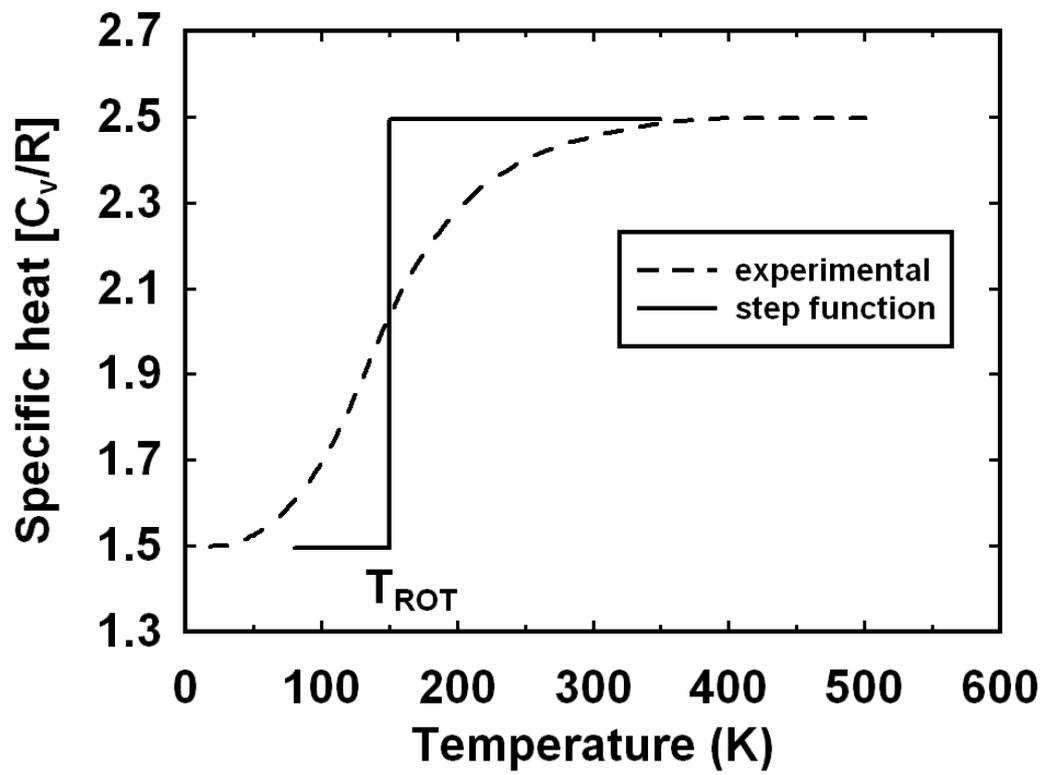

**Figure 1.** Approximation of the specific heat of hydrogen (After Zemansky [4]) by a step function.